\documentclass[11.5pt,twocolumn,letterpaper]{article}

\usepackage[final]{cvpr}      
\usepackage{graphicx}
\usepackage{amsmath}
\usepackage{amssymb}
\usepackage{booktabs}
\usepackage{bbold}
\usepackage{multirow}

\usepackage[pagebackref,breaklinks,colorlinks]{hyperref}

\usepackage[capitalize]{cleveref}
\crefname{section}{Sec.}{Secs.}
\Crefname{section}{Section}{Sections}
\Crefname{table}{Table}{Tables}
\crefname{table}{Tab.}{Tabs.}

\def\cvprPaperID{*****} 
\def\confName{CVPR}
\def\confYear{2022}

\begin{document}

\title{LightDefectNet: A Highly Compact Deep Anti-Aliased Attention Condenser Neural Network Architecture for Light Guide Plate Surface Defect Detection}

\author{Carol Xu$^{2,*}$, Mahmoud Famouri$^{2,*}$, Gautam Bathla$^{2,*}$, Mohammad Javad Shafiee$^{1,2*}$, and Alexander Wong$^{1,2*}$ \\
$^1$University of Waterloo, Waterloo, Ontario, Canada\\
$^2$DarwinAI,  Waterloo, Ontario, Canada \\
$^*$equal contribution
}

\maketitle

\begin{abstract}
Light guide plates are essential optical components widely used in a diverse range of applications ranging from medical lighting fixtures to back-lit TV displays.  An essential step in the manufacturing of light guide plates is the quality inspection of defects such as scratches, bright/dark spots, and impurities.  This is mainly done in industry through manual visual inspection for plate pattern irregularities, which is time-consuming and prone to human error and thus act as a significant barrier to high-throughput production.  Advances in deep learning-driven computer vision has led to the exploration of automated visual quality inspection of light guide plates to improve inspection consistency, accuracy, and efficiency.  However, given the cost constraints in visual inspection scenarios, the widespread adoption of deep learning-driven computer vision methods for inspecting light guide plates has been greatly limited due to high computational requirements.  In this study, we explore the utilization of machine-driven design exploration with computational and ``best-practices'' constraints as well as L$_1$ paired classification discrepancy loss to create LightDefectNet, a highly compact deep anti-aliased attention condenser neural network architecture tailored specifically for light guide plate surface defect detection in resource-constrained scenarios.  Experiments show that LightDetectNet achieves a detection accuracy of $\sim$98.2\% on the LGPSDD benchmark while having just 770K parameters ($\sim$33$\times$ and $\sim$6.9$\times$ lower than ResNet-50 and EfficientNet-B0, respectively) and $\sim$93M FLOPs ($\sim$88$\times$ and $\sim$8.4$\times$ lower than ResNet-50 and EfficientNet-B0, respectively) and $\sim$8.8$\times$ faster inference speed than EfficientNet-B0 on an embedded ARM processor.
\end{abstract}
\begin{figure}
    \centering
    \includegraphics[width=8cm]{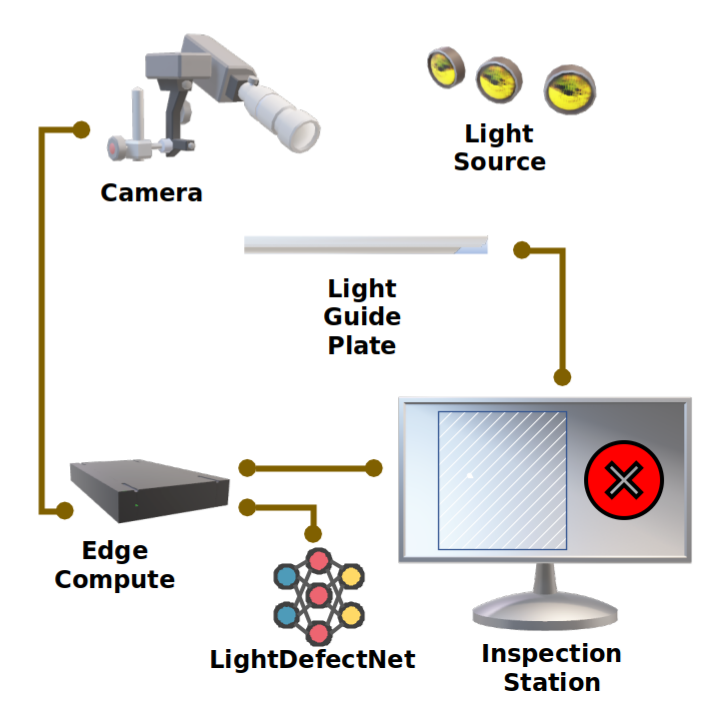}
    \caption{Light guide plate surface defect inspection workflow using LightDefectNet.  A light guide plate on an assembly line is imaged using a camera unit.  An edge compute unit equipped with LightDefectNet conducts defect detection, and an inspector may use the inspection station to examine defective light guide plates.}
    \label{fig:mainflow}
\end{figure}

\section{Introduction}
\label{sec:intro}
Light guide plates~\cite{feng2004high} are one of the main optical components used in different devices such as medical lighting fixtures and both commercial flat panels and back-lit displays. In the LED LCD panels as an example,  the light from the LED lamp passes through the light guide plate to be sprinkled on the surface of the large screen. The main role of the light guide plate is to  distribute the light evenly and to illuminate the surface. As such, it is very important to make sure there is no defect on the plate and visual inspection process plays a key role to identify possible defects.

Visual quality inspection of light guide plates is very challenging for a number of reasons;  the low contrast between the defect and the background, uneven brightness, and complex gradient texture make identifying the possible defects for an automated system tremendously difficult. As such, identifying possible defects on light guide plates are still  done mostly manually.

\begin{figure*}[ht]
    \centering
    \includegraphics[width=16cm]{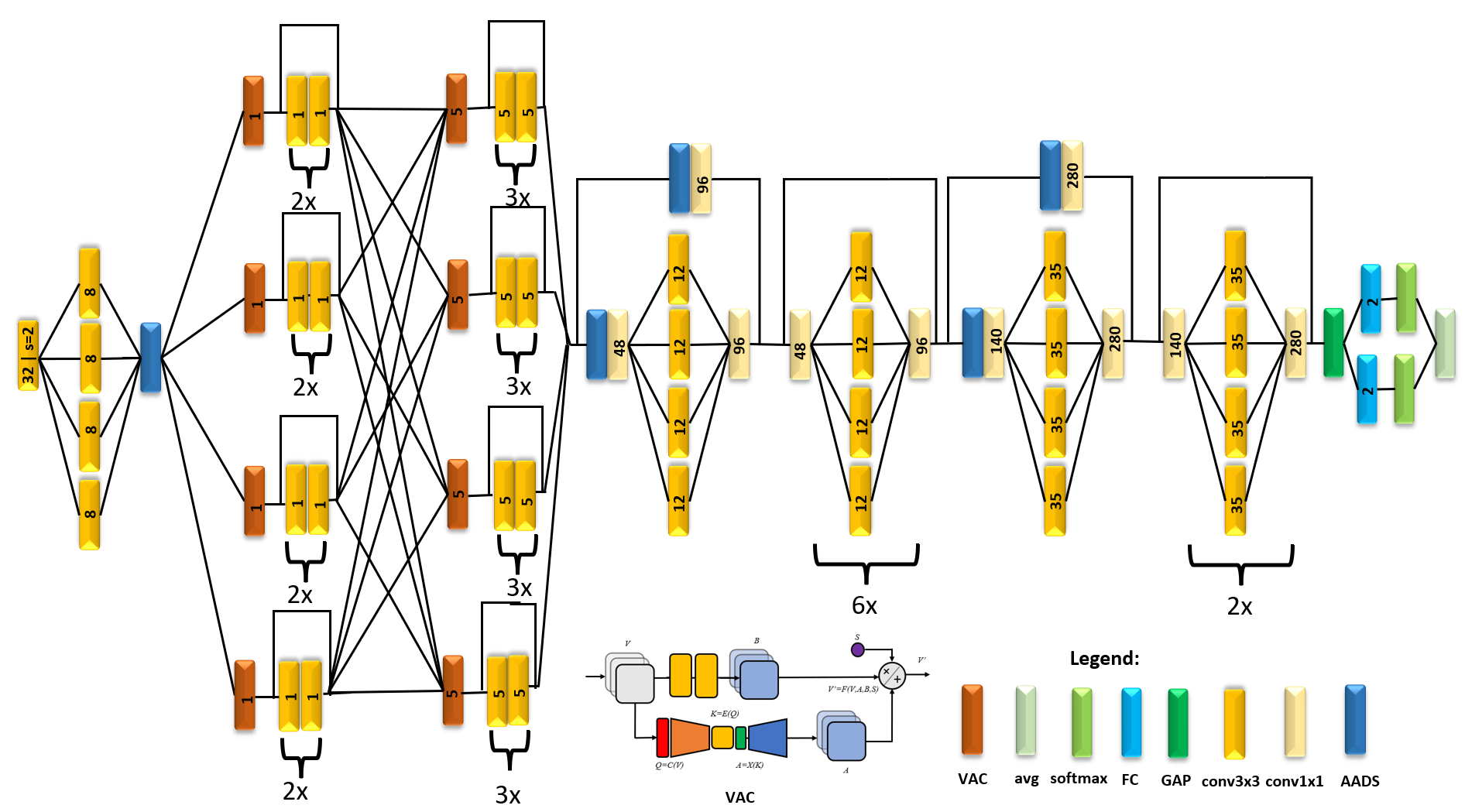}
    \caption{LightDefectNet architectural design.  It possesses a highly tailored attention condenser network architecture that was automatically identified using a machine-driven design exploration approach based on computational and ``best-practices'' constraints.}
    \label{fig:network}
\end{figure*}

Recently the promises of deep learning~\cite{robinson2020learning,he2016deep,reiss2021panda,bergman2020classification} has motivated the development of new high performing deep neural networks for different manufacturing tasks for the purpose of improving the automation including the inspection systems~\cite{li2021end,shafiee2021tinydefectnet,9306797}. However, this field of research is still in its infancy given the constraints these types of systems require to satisfy and the limitations need to be addressed,  including i) high efficiency requirements, ii) with high accuracy and robustness, and iii) limited training data samples of different defected. Building deep neural networks satisfying the aforementioned constraints is time-consuming and usually impossible for non-expert users.

In this study, we take advantage of a machine-driven design exploration approach to specifically address the first two challenges and to a certain degree the third challenge of lacking enough training data by effectively exploring the architecture design space to automatically identify a tailored attention condenser network architecture (which we name LightDefectNet) based on computational and ``best-practices'' constraints. The inspection workflow using LightDefectNet is shown in Figure~\ref{fig:mainflow}.

\section{Methodology}
In this study, we leverage the concept of Generative Synthesis~\cite{wong2018ferminets} to automatically identify the macro- and micro-architecture designs of the proposed LightDefectNet. More specifically, the design exploration process is formulated as a constrained optimization problem where the optimal network architecture is determined by  finding the optimal generator $G^\star(\cdot)$ which can generate network architectures $\{\mathcal{N}_s|s \in  S\}$ maximizing a universal performance function $U$ \cite{wong2019netscore} subject to a set of constraints:
\begin{align}
      G^{\star} = \underset{G^{'}} {\max} U \Big(G(s)\Big) \;\;\; \text{s.t.} \;\;\; \mathbb{1}_g(G(s)) = 1 \;\;\; \forall s \in S,
\end{align}
where $S$ is a set of seeds. The set of constraints are defined by a predefined set of operational requirements formulated via an indicator function $\mathbb{1}_g(\cdot)$.
The synthesis process is done within an iterative approach where at each step, the previous generator  $\bar{G}(\cdot)$ is evaluated by an inquisitor $I$ and based on its newly generated architectures $\mathcal{N}_s$. A new generator solution is evaluated based on the universal performance function $U$ by an indirect evaluation process.

 We define a residual design prototype~\cite{he2016deep} to initialize the design process. The indicator function $\mathbb{1}_g(\cdot)$ is formulated such that it accounts for a combination of computational and ``best-practices'' constraints driven by architecture design lessons learned over the years by the community: i) number of floating-point operations (FLOPs) is under 100M FLOPs for resource-constrained manufacturing scenarios, (ii) pointwise strided convolutions (first introduced in the original residual network design~\cite{he2016deep} and continues to be leveraged in the recent RegNet design~\cite{radosavovic2020designing}) are restricted, as their use can lead to considerable information loss within the network, (iii) downsampling can only be conducted after the input layer via antialiasing downsampling (AADS)~\cite{zhang2019making}, as they have been shown to significantly improve network stability and robustness.  Furthermore, we incorporate attention condensers~\cite{wong2020tinyspeech, wong2020attendnets} as a viable design pattern into the machine-driven design exploration process, which are highly efficient self-attention mechanisms that learns and produces a condensed embedding characterizing joint local and cross-channel activation relationships for the purpose of selective attention. How and where attention condensers are leveraged, along with the rest of the micro-architecture and macro-architecture designs of LightDefectNet is left to the machine-driven design exploration process to automatically determine how best to satisfy the  constraints.

\subsection{Network Architecture Design}
 The network architecture design of LightDefectNet is shown in Figure~\ref{fig:network}. A number of key observations can be made about the generated LightDefectNet architecture;
 \begin{enumerate}
     \item {\bf Early-stage self-attention:} it can be observed that visual attention condensers are leveraged heavily in the early stages of the network architecture.  The presence of visual attention condensers in these early stages enhance selective focus on important low-to-medium level visual indicators pertinent to light guide plate defects, while at the same time improving representational efficiency early on.
     \item \textbf{Heterogeneous columnar design:} a heterogeneous combination of columnar design patterns is exhibited in the proposed architecture design, with more independent columns with fewer stages of intermediate interaction in the earlier stages but more interactions in the later stages, which gives great balance between representational power and disentanglement with efficiency.  Furthermore, the fully-connected (FC) layer exhibits a columnar architecture where there are two softmax outputs which are then aggregated into the final softmax output to improve generalization and robustness particularly in low-data regimes.
     \item \textbf{Anti-aliased design:} a heavy use of anti-aliased downsampling (AADS) operations is exhibited across the architecture, which leads to improve robustness and stability compared to the use of traditional downsampling operations such as max-pooling. In this study, due to the columnar design of the FC layer, we leverage a loss function that maximizes the discrepancy between the soft outputs of the two FC columns to improve robustness and generalization of LightDefectNet in low-data regimes.
     \item \textbf{Heterogeneous design diversity:} heterogeneous microarchitecture and macroarchitecture design diversity is exhibited across the architecture due to the ability for the machine-driven design exploration strategy to determine the optimal microarchitecture designs in a fine-grained manner tailored around operational constraints and striking a strong balance between accuracy and efficiency.
 \end{enumerate}

\subsection{Results \& Discussion}
To explore the efficacy of the proposed LightDefectNet for light guide plates defect detection, we evaluated its performance on the LGPSDD (Light Guide Plate Surface Defect Detection) benchmark~\cite{9306797}.

\subsection{Benchmark Data} The LGPSDD (Light Guide Plate Surface Defect Detection) benchmark~\cite{9306797} used in this study consists of 822 images captured of light guide plates moving on a conveyor belt using an image acquisition platform that employs a line-scan camera and a multi-angle lighting source.  The light guide plates captured in the LGPSDD benchmark has high physical diversity in terms of light guide point density, light guide brightness, as well as defect morphology and size~\cite{9306797}. This resulted in 422 defective samples and 400 non-defective samples, with a training/test split of 25\%/75\% as described in~\cite{9306797} to better mimic the typical low annotated data scenario seen in manufacturing applications.  The images are 224 $\times$ 224 in size, and these same dimensions were used in the input dimensions for the tested neural network architecture designs in this study.

\begin{center}
    \begin{table}
    \center
        \caption{ Quantitative results of the proposed LightDefectNet architecture compared to other tested architectures.}
    \setlength{\tabcolsep}{0.1cm}
    \begin{tabular}{l || c c c c }
        \multirow{ 2}{*}{ Model} &  Acc  &  Param & FLOPs &  Inf. Speed  \\
        &(\%) & (M) & (M) & (ms)\\
        \hline \hline
         ResNet-50 & 92.8 & 25.6 & 8200 & 83\\
        \hline
         EfficientNet-B0 & 98.0 & 5.3 & 780 & 88\\
        \hline
         MnasNet & 89.4 & 3.9 & 630 & 89\\
        \hline
         MobileNetV3 (Large) & 97.8 & 5.4 & 438 & 56\\
        \hline
        LightDefectNet & \textbf{98.2} & \textbf{0.77} & \textbf{93} & \textbf{10}
    \end{tabular}
    \label{tab:res}
    \end{table}
\end{center}

\subsection{Evaluated Architectures} In this study, in addition to the proposed LightDefectNet, we evaluated the performance of the ResNet-50~\cite{he2016deep} network architecture on the same LGPSSD benchmark for reference purposes, as well as several state-of-the-art efficient deep neural network architectures, including EfficientNet-B0~\cite{tan2020efficientnet}, MobileNetV3~\cite{howard2019mobilenet}, and MnasNet~\cite{tan2019MnasNet}. All architecture designs and experiments were conducted within the Pytorch deep learning framework.  In addition to quantitatively assessing the accuracy of each deep neural network architecture, we also evaluated the architectural complexity, theoretical computational complexity, as well as inference speed on an embedded ARM v8.2 64-bit 2.26GHz processor.

\subsection{Training Policies} All tested network architectures were trained using stochastic gradient descent optimization for 100 epochs and batch size of 5.  Different learning rates were used for optimal performance based on empirical analysis: $5.0\times10^{-5}$ for ResNet-50, $1.0\times10^{-3}$ for EfficientNet-B0 and LightDefectNet, $1.3 \times 10^{-3}$ for MobileNetV3, and $3.1\times10^{-4}$ for MnasNet.

\subsection{Results}

\textbf{Quantitative performance and complexity}: Table~\ref{tab:res} illustrates the quantitative performance, architectural complexity, computational complexity, and inference speed of the proposed LightDefectNet architecture compared to the ResNet-50 architecture as well as several state-of-the-art efficient architectures. A number of observations can be made from the quantitative results.  First of all, it can be observed that the proposed LightDefectNet architecture consists of just $\sim$770K parameters, which is significantly smaller than ResNet-50 as well as the competing state-of-the-art efficient architectures.  More specifically, LightDefectNet is $\sim$33$\times$ smaller compared to the ResNet-50 architecture while achieving higher accuracy, and $\sim$6.9$\times$ smaller compared to the highly efficient EfficientNet-B0 (the most accurate architecture outside of LightDefectNet). Second, in terms of computational complexity, the proposed LightDefectNet architecture requires only $\sim$93M FLOPs, which is significantly lower than ResNet-50 as well as the tested state-of-the-art efficient architectures.  More specifically, LightDefectNet requires $\sim$88$\times$ fewer FLOPs compared to the ResNet-50 architecture, and $\sim$8.4$\times$ fewer FLOPs compared to EfficientNet. Third, from an accuracy perspective, LightDefectNet achieved the highest accuracy amongst  the tested architectures.  These results illustrate the strong balance achieved by the proposed LightDefectNet in terms of accuracy, architectural complexity, and computational complexity, making it very well-suited for high-performance light guide plate defect detection in resource-constrained manufacturing environments.

\textbf{Embedded inference speed}: We further explore real-world operational efficiency of the proposed LightDefectNet architecture in embedded scenarios by evaluating its run-time latency (at a batch size of 10) on an embedded ARM v8.2 64-bit 2.26GHz processor and compared with the other tested architectures.  It can be observed from Table~\ref{tab:res} that the proposed LightDefectNet architecture achieves a runtime latency of 10~ms per sample, which is significantly lower than ResNet-50 as well as the tested state-of-the-art efficient architectures.  More specifically, LightDefectNet is 8.3$\times$ faster when compared to the ResNet-50 architecture, 8.8$\times$ faster when compared to the EfficientNet-B0 architecture, and 5.6$\times$ faster when compared to the MobileNetV3 architecture (the fastest architecture outside of LightDefectNet).  The significant speed advantages of the proposed LightDefectNet architecture when compared to the other tested architectures make it very well-suited for use on embedded edge compute devices for high-throughput manufacturing scenarios, as well as illustrate the effectiveness of utilizing a machine-driven design exploration strategy with ``best-practices'' constraints for creating highly customized network architectures tailored specifically for industrial tasks in manufacturing scenarios.

\section{Conclusions}
Here, we conducted an exploration into utilizing machine-driven design exploration with computational and ``best-practices`` constraints as well as L$_1$ paired classification discrepancy loss for the creation of highly compact deep neural network architectures for the task of light guide plate surface defect detection.  Experimental results demonstrated that the proposed LightDefectNet was able to achieve a detection accuracy of $\sim$98.2\% on the LGPSDD benchmark while possessing significantly reduced architectural and computational complexity when compared to state-of-the-art efficient deep neural network architectures.  Furthermore, we demonstrated that the proposed LightDefectNet achieves significantly faster inference speed on an embedded ARM processor, making it very well-suited for light guide plate defect detection in high-throughput, resource-constrained manufacturing scenarios.  As a future direction, we aim to further explore the leveraging of this machine-driven design exploration strategy for producing highly efficient yet high-performing deep neural network architectures for other critical manufacturing applications as well as for other sensing modalities such as acoustic sensors for predictive maintenance.

{\small
\bibliographystyle{ieee_fullname}
\bibliography{egbib}
}

\end{document}